\documentclass[a4paper,11pt]{article}
\usepackage{pos}
\usepackage{amsmath}
\usepackage{orcidlink}
\usepackage{graphicx}
\usepackage{hyperref}
\usepackage{microtype}
\usepackage{verbatim}

\newcommand{\llangle}{\langle \kern-.17em \langle}
\newcommand{\rrangle}{\rangle \kern-.17em \rangle}

\newcommand{\orcidauthorMASON}{0000-0002-1857-1085}
\newcommand{\orcidauthorBENNETT}{0000-0002-1678-6701}
\newcommand{\orcidauthorLUCINI}{0000-0001-8974-8266}
\newcommand{\orcidauthorPIAI}{0000-0002-2251-0111} 
\newcommand{\orcidauthorRINALDI}{0000-0003-4134-809X} 
\newcommand{\orcidauthorVADACCHINO}{0000-0002-5783-5602}
\newcommand{\orcidauthorZIERLER}{0000-0002-8670-4054}

\newcommand{\beq}{\begin{equation}}
\newcommand{\eeq}{\end{equation}}
\newcommand{\beqs}{\begin{eqnarray}}
\newcommand{\eeqs}{\end{eqnarray}}

\title{Finite-temperature Sp(4) Yang-Mills theory: towards the continuum}

\author*[a,b,c]{Fabian Zierler\,\orcidlink{\orcidauthorZIERLER}}
\author[b,d]{Ed Bennett\,\orcidlink{\orcidauthorBENNETT}}
\author[e,d,f]{Biagio Lucini\,\orcidlink{\orcidauthorLUCINI}}
\author[g]{David Mason\,\orcidlink{\orcidauthorMASON}}
\author[b,c]{Maurizio Piai\,\orcidlink{\orcidauthorPIAI}}
\author[e,h]{Enrico Rinaldi\,\orcidlink{\orcidauthorRINALDI}}
\author[i]{Davide Vadacchino\,\orcidlink{\orcidauthorVADACCHINO}}

\affiliation[a]{Technical University of Munich, TUM School of Natural Sciences, Physics Department, James-Franck-Str.~1, 85748 Garching bei München, Germany}
\affiliation[b]{Centre for Quantum Fields and Gravity, Faculty  of Science and Engineering, Swansea University, Singleton Park, SA2 8PP, Swansea, United Kingdom}
\affiliation[c]{Department of Physics, Faculty of Science and Engineering, Swansea University (Park Campus), Singleton Park, SA2 8PP Swansea, Wales, United Kingdom}
\affiliation[d]{Swansea Academy of Advanced Computing, Swansea University (Bay Campus), Fabian Way, SA1 8EN Swansea, Wales, United Kingdom}
\affiliation[e]{School of Mathematical Sciences, Queen Mary University of London, Mile End Road, London, E1 4NS, United Kingdom}
\affiliation[f]{Department of Mathematics, Faculty of Science and Engineering, Swansea University (Bay Campus), Fabian Way, SA1 8EN Swansea, Wales, United Kingdom}
\affiliation[g]{Jülich Supercomputing Centre, Forschungszentrum Jülich, D-52425 Jülich, Germany}
\affiliation[h]{RIKEN Center for Quantum Computing, RIKEN, 2-1 Hirosawa, Wako, Saitama, 351-0198, Japan}
\affiliation[i]{Centre for Mathematical Science, University of Plymouth, Plymouth, PL4 8AA, United Kingdom}
\emailAdd{fabian.zierler@tum.de}

\abstract{
\begin{center}
\href{https://telos-collaboration.github.io}{ \includegraphics[height=1cm]{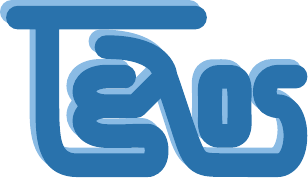}}\\
(on behalf of the TELOS collaboration)
\end{center}

We present numerical results obtained in a finite-temperature study of the Sp(4) Yang-Mills theory on the lattice. We study  its first-order confinement/deconfinement phase transition, by reconstructing the density of states via the Logarithmic Linear Relaxation (LLR) algorithm. We perform our measurements on lattices with different  extents of space and time (and aspect ratios). We estimate the size of discretisation and finite-volume  artefacts. We find clear  signatures of a first-order transition. We determine the critical coupling, the specific heat, and  the surface tension, for finite extents of the thermal circle, and use the results to set bounds for the continuum theory.}

\FullConference{The 42nd International Symposium on Lattice Field Theory (LATTICE2025)\\
2-8 November 2025\\
Tata Institute of Fundamental Research, Mumbai, India\\}

\begin{document}

\begin{flushright}
\textsc{TUM-EFT 207/26}
\end{flushright}

\maketitle

\section{Introduction}\label{sec:intro}

Non-Abelian, confining gauge theories, possibly coupled to new fermion matter field content, are promising candidates for the short-distance completion of proposals for  Beyond the Standard Model (BSM) physics with composite dynamical origin. Their properties can be used to address a broad range of open phenomenological questions, ranging from the origin of dark matter, to electroweak and Higgs physics, and  to the generation of mass of SM fermions---see for instance the reviews in Refs.~\cite{Panico:2015jxa,Cacciapaglia:2020kgq,Cirelli:2024ssz,Bennett:2023wjw} and references therein. One of the exciting aspects of some of these theories is that they lead to a relic stochastic background of  gravitational waves (GWs), if the new physics underwent a phase transition in the early universe~\cite{Witten:1984rs}. This signal is potentially detectable by the next generation of ground and space based experiments~\cite{ET:2019dnz,Caprini:2019egz}. Yet, computing a precise prediction of  the gravitational wave power spectrum is not straightforward.   To determine the gravitational wave spectrum within the thin-wall approximation, one requires knowledge of the latent heat and the confined-deconfined surface tension~\cite{Huang:2020crf,Halverson:2020xpg,Kang:2021epo}, which are the subjects of the $SU(N)$ lattice studies in Refs.~\cite{Lucini:2002ku,Borsanyi:2022xml,Rindlisbacher:2025dqw}. Alternative approaches based on models involving the Polyakov loop have also been proposed---see for instance Refs.~\cite{Huang:2020crf,Reichert:2021cvs,Pasechnik:2023hwv}.

The strongly coupled  nature of  non-Abelian gauge theories requires non-perturbative tools in order to make theoretical predictions. The methods of lattice field theory provide a first-principles, systematically improvable approach to  determine numerically the non-perturbative properties of these theories. They rely on sampling the space of possible gauge field configurations, $\{U\}$, to determine the expectation value, or ensemble average, of operators of interest, $O$, defined as
\begin{align}
    \label{eq:expectation_value}
    \langle O \rangle =&\frac{1}{Z} \int D[U] \mathcal{O}[U] \exp \left( - \beta S[U] \right)\,,
\end{align}
where the normalisation of the ensemble average is given by
\begin{align}
    \label{eq:partition_function}
     Z =& \int D[U] \exp \left( - \beta S[U] \right).
\end{align}
The existence of  first-order phase transitions in Yang-Mills theories has been established numerically  for gauge theories with  $SU(N>2)$~\cite{Lucini:2002ku,Panero:2009tv,Lucini:2012wq,Rindlisbacher:2025dqw}, $Sp(2N>2)$~\cite{Holland:2003kg}, and $G_2$~\cite{Cossu:2007dk,Pepe:2006er} gauge group.
It is challenging to achieve the required precision  with standard lattice algorithms based on importance sampling, as their effectiveness is limited in the presence of  a first-order phase transition.
Due to the phenomenon of phase coexistence, the distribution in Eqs.~\eqref{eq:expectation_value} and \eqref{eq:partition_function} has a bimodal character in the proximity of the transition, and standard importance sampling algorithms struggle to sample the path integral efficiently, because the Markov chains tend to get stuck in one phase.
 
The Logarithmic Linear Relaxation (LLR) algorithm~\cite{Langfeld:2012ah,Langfeld:2015fua} provides an alternative algorithm to describe Yang-Mills theories in proximity of a first-order phase transition. Building  on our earlier work on $SU(3)$ on a fixed lattice volume~\cite{Lucini:2023irm}, and on the thermodynamic limit $Sp(4)$ for fixed thermal circle, $N_t=4$~\cite{Bennett:2024bhy}, in this contribution we present  new results for $Sp(4)$ with $N_t=5$. This research is a necessary step towards the continuum limit. More details can be found in Ref.~\cite{Bennett:2025whm}. 

\section{Density-of-states \& Logarithmic Linear Relaxation }\label{sec:llr}

We circumvent the limitations of importance sampling methods by using an approach based on the density of states, $\rho(E)$. For operators that depend only on the energy, $\mathcal O(S[U]=E)$, we rewrite Eqs.~\eqref{eq:expectation_value} and \eqref{eq:partition_function} as a one dimensional integral over the energy 
\begin{align}
    \langle O \rangle &= \frac{1}{Z} \int {\rm d}E \rho(E) \mathcal{O}(E) \exp \left(- \beta S[E] \right)\,,\\
    Z &= \int {\rm d}E \rho(E) \exp \left(-\beta S[U] \right)\,,
\end{align}
where we have defined the density of states as
\begin{align}
    \rho (E) &= \int D[U] \delta\left(S[U] - E \right)\,.
\end{align}
In order to determine the density of states precisely and reliably, we follow the process outlined in Ref.~\cite{Langfeld:2012ah,Langfeld:2015fua}, and
 employ a linear logarithmic approximation of the density of states. We write
\begin{align}
    \rho(E) = \rho(0) \exp\left[ - \int_0^E {\rm d}\bar E a(\bar E) \right]\,,
\end{align}
where $a(E)$ is a piecewise-linear function, defined by splitting the energy into small intervals. The coefficients $a_n$ and $c_n$ determine the density of states as
\begin{align}
    \log \rho(E) \simeq a_{n} \left( E^0_{n} - E \right) + c_{n}\,,
\end{align}
in a given interval centred around $E^0_{n}$. The coefficients $c_{n}$ are determined from requiring continuity---up to the first coefficient, $c_0$, which corresponds to $\rho(0)$ and drops out of the ensemble averages. 

The remaining coefficients, $a_{n}$, are determined by considering the following expectation value, which we denote with double brackets:
\begin{align}\label{eq:double_bracket}
    \llangle O \rrangle(\hat a) &= \int D[A] O[A] W\!(E_n,\delta) \exp\left[ \left(  S[A] - E_n \right) \hat a \right]\,,
    \end{align}
    and where we introduced the function
    \begin{align}
    W(x,\delta) &= \begin{cases} 1 &\text{if } x \in \left[ - \delta/2, + \delta/2 \right] \\ 0 &\text{else} \end{cases}\,,
\end{align}
in which $\hat a$ is for now an arbitrary parameter. The purpose of $W$ is to restrict the energy range to a finite interval. We compute $a_n$ by setting $O = S[A] - E_n$ and requiring its vanishing 
\begin{align}\label{eq:db_energy}
    \llangle O = S[A] - E_n \rrangle(\hat a = a_n) = 0\,.
\end{align}

To find the (unique) root of Eq.~(\ref{eq:db_energy}), and hence $a_n$, we define a stochastic, non-linear, iterative process, with the following prescription for the $(i+1)$-th iteration step
\begin{align}
    a_{n}^{(i+1)} = a_n^{(i)} - b^{(i)} \llangle S[A] - E_n \rrangle (a_n^{(i)}).
\end{align}
Initially, we start with several Newton-Raphson update steps, for which $b_{i}\equiv 1$. Then we adopt the simple choice $b_i = 1/i$ for the Robbins-Monro prescription,  that satisfy the relations
\begin{align}
    \sum_i b_i \to \infty\,, \quad\quad\quad\quad \sum_i b_i^2 = \text{finite}\,.
\end{align}

This procedure may be affected by a potential ergodicity problem. By constraining every Markov chain to a single energy interval when calculating $\llangle O \rrangle$, we might miss gauge configurations with energies in the same interval that are only connected via intermediate energies outside the selected interval.
To circumvent this potential issue, we implement additionally a \textit{parallel tempering} approach. We consider energies in the range $\left[E_{\min} , E_{\max} \right]$ and choose $N_{\rm rep}$ evenly-spaced, overlapping intervals and start a Markov chain for every energy interval. After every update we swap the Markov chains of two neighbouring intervals, $n$ and $n+1$, with a probability of
\begin{align}
    P_{\rm swap} = \min\left( 1, e^{\left( S[A^{(n)}] - S[A^{(n+1)}] \right) \left( a^{(n)} - a^{(n+1)} \right)} \right).
\end{align}
We allow the first and last interval to probe energies below $E_{\min}$ and above $E_{\max}$, respectively.

\section{Lattice Setup}\label{sec:setup}
We express the Wilson gauge action in terms of the plaquette, $U=U_{\mu\nu}$, as
\begin{align} 
    S[U] = \frac{1}{N_c} \sum_x \sum_{\mu < \nu} {\rm Re}~{\rm Tr} \left[ 1 - U_{\mu\nu}(x) \right] \equiv 6N_t N_s^3 \left( 1 - u_p[U] \right)\,,
\end{align}
where $N_c=2N=4$ for $Sp(4)$.
The hypercubic lattice has lattice spacing $a$ and $N_t$ lattice sites in the temporal (thermal) direction and $N_s$ lattice sites in the space directions. The averaged plaquette, $u_p$, will be later  used as a proxy for the energy, $E$. We impose periodic boundary conditions, and interpret the temporal lattice extent (with $N_t < N_s$) in terms of the equilibrium temperature. We use the HiRep code \cite{HiRepSUN, DelDebbio:2008zf} extended to symplectic gauge groups \cite{HiRepSpN} with support for the LLR based on restricted heatbath updates with domain decomposition and over-relaxation \cite{mason_HiRep_LLR_v1.0.0,Lucini:2023irm}. 

\begin{table}
    \centering
    \caption{Lattice parameters controlling the LLR algorithm for the $Sp(4)$ Yang-Mills theory.}
    \begin{tabular}{|c|c|c|c|c|c|c|c|} \hline
$N_t$ & $N_s$ & $u_{p}^{\rm min}$ & $u_{p}^{\rm max}$ & $N_{\rm rep}$ & $N_{\rm repeats}$ & $n_{\rm NR}$ & $n_{\rm RM}$ \\ \hline \hline 
5 & 48 & 0.588 & 0.592 & 48 & 25 & 10 & 60 \\
5 & 48 & 0.588 & 0.592 & 96 & 25 & 10 & 50 \\
5 & 56 & 0.588 & 0.592 & 128 & 25 & 10 & 50 \\
5 & 56 & 0.588 & 0.592 & 48 & 25 & 10 & 50 \\
5 & 56 & 0.588 & 0.592 & 96 & 25 & 10 & 50 \\
5 & 64 & 0.588 & 0.592 & 95 & 20 & 7 & 50 \\
5 & 72 & 0.588 & 0.592 & 95 & 20 & 11 & 50 \\
5 & 80 & 0.588 & 0.59 & 64 & 20 & 15 & 30 \\
\hline \hline 
\end{tabular}

    \label{tab:runs}
\end{table}
We show the value of the lattice parameters used for the LLR study  in table~\ref{tab:runs}.
The energy range corresponds to values of the average plaquette in the interval $\left[ u_p^{\min}, u_p^{\max} \right]$, divided in $N_{\rm rep}$ small intervals. We use $n_{\rm NR}$ Newton-Raphson updates followed by $n_{\rm RM}$ Robbins-Monro updates. In evaluating the double bracket expectation value in Eq.~\eqref{eq:double_bracket}, we first perform $300$ thermalisation steps, followed by $700$ measurement steps. 
We estimate the overall uncertainties by repeating the iteration procedure $N_{\rm repeat}$ times and performing a jackknife resampling analysis among all repetitions. 

\section{Observables}

We first consider the probability distribution for a given energy, $E$, defined as
\begin{align}
    P_\beta(E) = \frac{1}{Z(\beta)} \rho(E) e^{-\beta E}\,.
\end{align}
At the critical coupling, $\beta_c$, the two phases  coexist and the distribution must show two peaks. At  finite spatial volume, we can define the critical coupling by dialling it until the peaks have equal heights. The density of states does not depend on $\beta$, and, once we have obtained $\rho(E)$, this tuning can determine $\beta_c$ very accurately.

We can further use the probability density to estimate the interface tension from the height of the peaks of the distribution relative to the height of an intermediate plateau arising in the interval between them, due to the presence of an interface. We estimate this effect by considering the minimal value of $P_\beta$ between the two maxima, which in the thermodynamic limit takes the form
\begin{align} 
     \frac{P_{\rm min}}{P_{\rm max}} &\propto \sqrt{N_s} \exp \left( - 2 \frac{N_s^2}{N_t^2} \frac{\sigma_{cd}}{T_c^3}  \right)\,.
\end{align}
We measure the quantity 
\begin{align}
\label{eq:surface_tension}
\tilde I &\equiv-\frac{1}{2} \left(\frac{N_t}{N_s}\right)^2\log\left(\frac{P_{\rm min}}{P_{\rm max}}\right) + \frac{1}{4} \left(\frac{N_t}{N_s}\right)^2\log (N_s)\,,
\end{align}
which  in the thermodynamic limit satisfies $\lim_{N_s/N_t\to\infty} \tilde I  = \frac{\sigma_{cd}}{T_c^3}$.

Alternative definitions of $\beta_c$ are provided by the maximum of the specific heat, $C_V(\beta)$, and the minimum of the Binder cumulant, $B_V(\beta)$, respectively:
\begin{align}
    \label{eq:specific_heat}
    C_V(\beta) \equiv \frac{6V}{a^4} \left[ \langle u_p^2 \rangle_\beta - \langle u_p \rangle_\beta^2 \right],
    \quad \quad \quad \quad
    B_V(\beta) \equiv 1 - \frac{\langle u_p^4 \rangle_\beta}{3\langle u_p^2 \rangle_\beta^2}.
\end{align}
While these determinations of $\beta_c$ are guaranteed to coincide in the thermodynamic limit, this is not the case at finite volume. 

\section{Results}\label{sec:results}
\begin{figure}
    \centering
    \includegraphics[width=0.4\linewidth]{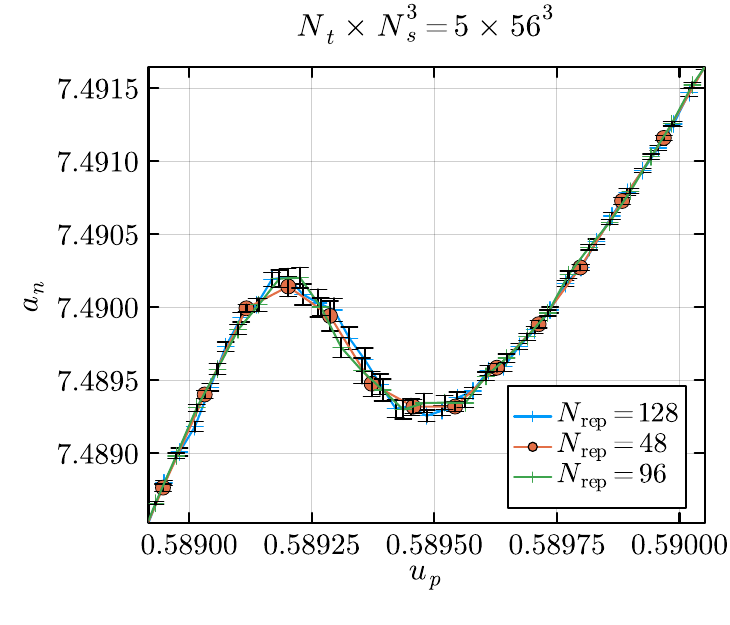} \hfill
    \includegraphics[width=0.4\linewidth]{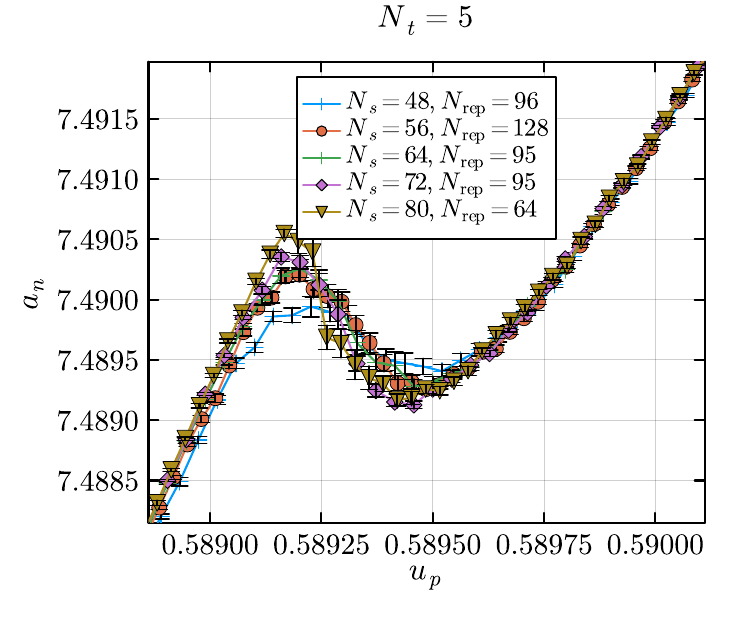}
    \caption{Left panel: final result for $a_n$ as a function of the average plaquette, $u_p$, evaluated at the centre of the intervals, for fixed choice of lattice volume, and by varying the number of energy intervals. Right panel: final results for $a_n$ as a function of the average plaquette, $u_p$, evaluated at the centre of the intervals, for all spatial volumes considered, with fixed $N_t=5$.}
    \label{fig:an_plots}
\end{figure}
\begin{figure}
    \centering
    \includegraphics[width=0.43\linewidth]{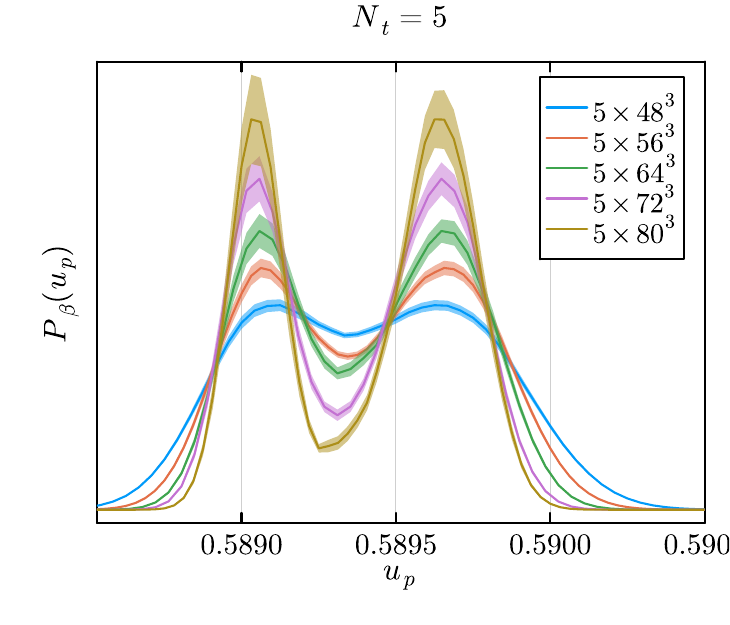} \hfill
    \includegraphics[width=0.40\linewidth]{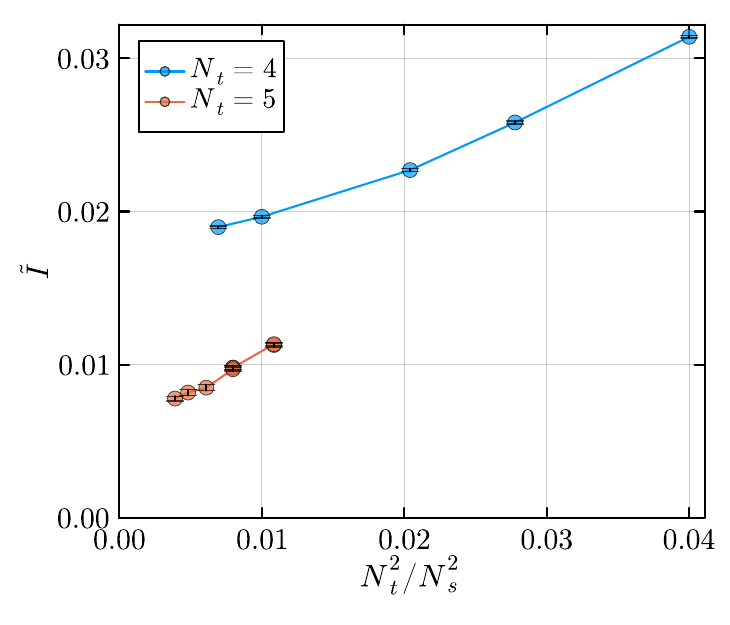}
    \caption{Left panel: probability distribution, $P_\beta(u_p)$, against the  central value of the average plaquette. The inverse gauge coupling, $\beta$, has been tuned such that the maxima are of equal height. Right panel:  surface tension term, $\tilde I$, extracted at finite volume, according to Eq.~\eqref{eq:surface_tension}.}
    \label{fig:distribution_interface}
\end{figure}
\begin{figure}
    \centering
    \includegraphics[width=0.33\linewidth]{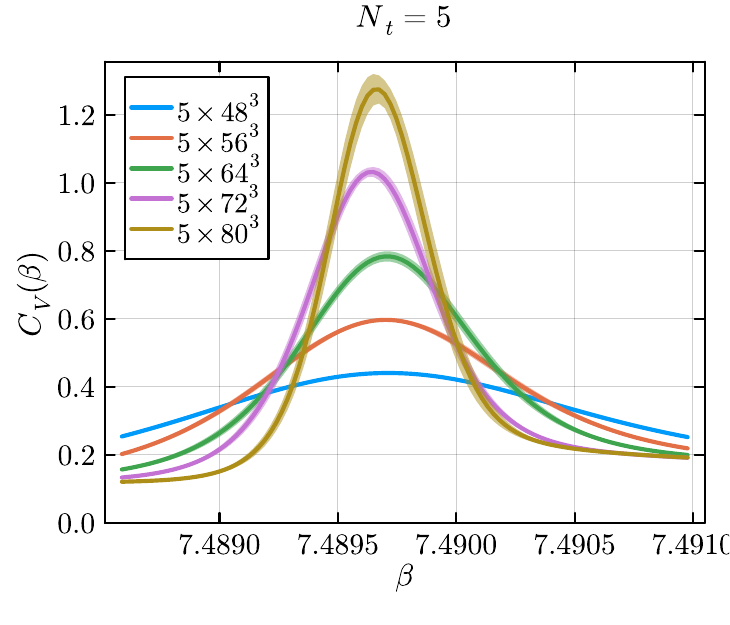}
    \includegraphics[width=0.33\linewidth]{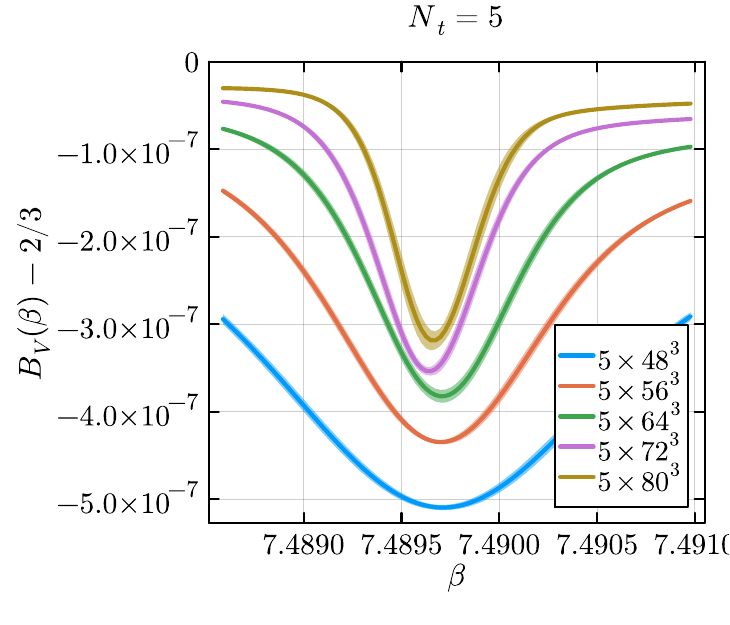}
    \includegraphics[width=0.32\linewidth]{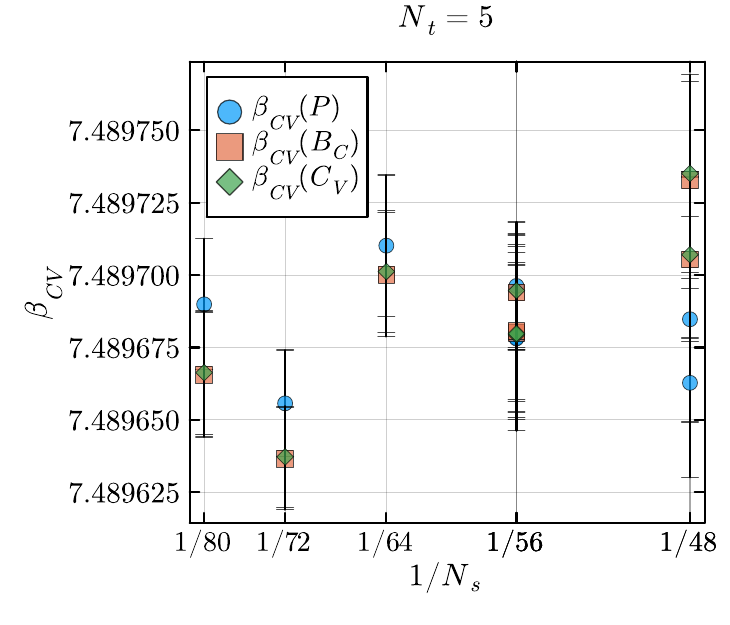}
    \caption{Left panel: specific heat, $C_V$, as a function of $\beta$. Middle panel: Binder cumulant, $B_V$, as a function of $\beta$.  In both case, the position of the extrema yields a determination of the critical coupling. Right panel: comparison between inverse critical coupling, $\beta_c$, as determined from the plaquette distribution ($P$), the specific heat ($C_V$), and the Binder cumulant ($B_V$).}
    \label{fig:cumulants}
\end{figure}

On the left-hand-side in Fig.~\ref{fig:an_plots}, we show the final values of $a_n$ plotted again the central value of the average plaquette, $u_p$. We observe excellent agreement between results obtained with the same lattice parameters, when varying the number of energy intervals. From this, we conclude that the chosen interval widths are sufficiently small as to not induce sizeable systematic artefacts.   

On the right panel of Fig.~\ref{fig:an_plots}, we compare the final values obtained for $a_n,$ for different choices of lattice volume. Starting at an aspect ratio of $N_s/N_t = 9.6$ we start to observe $a_n$ being multi-valued in $u_p$,  indicating the presence a first order-transition. This behaviour becomes more pronounced as the spatial volume (and aspect ratio) is increased. We observe that the required aspect ratios are larger than the values that were sufficient for $N_t=4$~\cite{Bennett:2024bhy}. Consequently, the computational cost for finer lattices grows substantially.

In Fig.~\ref{fig:distribution_interface}, in the left panel, we show the plaquette probability distribution tuned to the critical coupling for all available spatial volumes. A separation develops between the two peaks as the volume increases. Unfortunately, as in the case of $N_t=4$~\cite{Lucini:2023irm}, we do not observe the formation of a clear plateau in the interstice between peaks. In the right panel of the same figure, we show an estimate of the interface tension, $\tilde{I}$, defined in Eq.~\eqref{eq:surface_tension}, for both $N_t=4$ (based on the previously published numerical results) and $N_t=5$. We observe that $\tilde I$ for $N_t=5$ is approximately half of the results obtained at $N_t=4$. This finding is indicative of the presence of discretisation artefacts. This result should be assessed with the caveat that our analysis uses  the local minimum of the probability distribution a proxy for the eventual plateau. In the future, it might be helpful to address this point to consider the approach used in Ref.~\cite{Rindlisbacher:2025dqw} and elongate one of the spatial dimensions to provide a preferred direction for the formation of an interface.

In Fig.~\ref{fig:cumulants}, we show both the specific heat (left panel) and the Binder cumulant (middle), defined in Eqs.~\eqref{eq:specific_heat}. We observe the clear formation of local extrema, allowing a determination of the critical coupling. When comparing the resulting estimates of $\beta_c$ with the ones obtained from the plaquette distribution, we find agreement (within errors) for all volumes (right panel). Additionally, we observe only a moderate volume dependence. This  conclusion is in stark contrast with Ref.~\cite{Lucini:2023irm}, that reported sizeable volume dependence as well as a tension between different methodologies. This is interpreted as an encouraging partial result, in the approach to the continuum.

We close this report by commenting on early trends emerging from our analysis of ensembles with $N_t=6$, which form the stating point of a future project.  We have been so far unable to identify an energy range exhibiting clear visible signal of a first-order transition, for lattice volumes up to $6\times 72^3$, corresponding to aspect ratios larger than the ones needed for $N_t=4,5$. This finding is in line with the trend observed in transitioning between $N_t=4$ and $N_t=5$.

\section{Summary}

We have determined several observable quantities associated with the thermal first-order transition characterising deconfinement for the  $Sp(4)$ Yang-Mills theory. We considered  ensembles with  $N_t=5$, which extend and complement previous calculations performed with $N_t=4$. This serves as a step towards the continuum extrapolation. We demonstrated that finite volume effects are controllable, and that the systematics related to the methodology are reduced, in respect to $N_t=4$. Overall, the picture of a comparatively weak, first-order phase transition emerges. 

\acknowledgments
We thank Stephan J. Huber, David Mateos, and Manuel Reichert for useful discussions, and Frederic D. R. Bonnet for help in benchmarking our code.

E.B. and B.L. were supported by the EPSRC ExCALIBUR programme ExaTEPP (project EP/X017168/1). 
E.B. was supported by the STFC Research Software Engineering Fellowship EP/V052489/1.
E.B., B.L., M.P. and F.Z. were supported by the STFC Consolidated Grant No. ST/X000648/1.
B.L. was supported by the STFC Consolidated Grant No. ST/X00063X/1. 
B.L. and M.P. were supported by the STFC Consolidated Grant No. ST/T000813/1 and received funding from the ERC under the European Union’s Horizon 2020 research and innovation program under Grant Agreement No. 813942. 
D.M. was supported by a studentship awarded by the Data Intensive Centre for Doctoral Training, funded by the STFC grant ST/P006779/1. 
D.V. was supported by the STFC Consolidated Grant No. ST/X000680/1. 
F.Z. was supported by the Advanced ERC grant ERC-2023-ADG-Project EFT-XYZ.

{\bf High performance computing}---This work used the DiRAC Data Intensive service (CSD3) at the University of Cambridge,  the DiRAC Data Intensive service (DIaL3) at the University of Leicester and the DiRAC Extreme Scaling service (Tursa) at the University of Edinburgh, managed respectively by the University of Cambridge University Information Services, the University of Leicester Research Computing Service and by EPCC on behalf of the STFC DiRAC HPC Facility (www.dirac.ac.uk). The DiRAC service at Cambridge, Leicester, and Edinburgh are funded by BEIS, UKRI and STFC capital funding and STFC operations grants. DiRAC is part of the UKRI Digital Research Infrastructure. This work was supported by the Supercomputer Fugaku Start-up Utilization Program of RIKEN. This work used computational resources of the supercomputer Fugaku provided by RIKEN through the HPCI System Research Project (Project ID: hp230397). Numerical simulations have been performed on the Swansea SUNBIRD cluster (part of the Supercomputing Wales project). The Swansea SUNBIRD system is part funded by the European Regional Development Fund (ERDF) via Welsh Government.

{\bf Research Software Availability statement}---The results presented in this contribution are expanded in Ref.~\cite{Bennett:2025whm}. The workflow used to analyse these data is available at Ref.~\cite{analysis_release}. The modified HiRep code with support for the LLR is available at \cite{mason_HiRep_LLR_v1.1.0}. It is based upon \cite{HiRepSUN,HiRepSpN}.

{\bf Research Data Availability Statement}---The raw data generated in support of this work, and processed data derived from it, are available in machine-readable format at Ref.~\cite{data_release}. See also Ref.~\cite{Bennett:2025neg}, for a description of our approach to reproducibility and open science.

\bibliographystyle{jhep}
\bibliography{references}

@misc{HiRepSUN,
	author = {},
	title = {{G}it{H}ub - claudiopica/{H}i{R}ep: {H}i{R}ep repository --- github.com},
	howpublished = {\url{https://github.com/claudiopica/HiRep}},
	year = {},
}

@misc{HiRepSpN,
	author = {},
	title = {{G}it{H}ub - sa2c/{H}i{R}ep: {H}i{R}ep repository --- github.com},
	howpublished = {\url{https://github.com/sa2c/HiRep}},
	year = {},
}

@misc{data_release,
  author       = {Bennett, Ed and
                  Lucini, Biagio and
                  Mason, David and
                  Piai, Maurizio and
                  Rinaldi, Enrico and
                  Vadacchino, Davide and
                  Zierler, Fabian},
  title        = {Data release -- Finite-temperature Yang-Mills theories with the density of states method: towards the continuum limit},
  year         = 2025,
  publisher    = {Zenodo},
  version      = {1.0.0},
  doi          = {\href{{https://doi.org/10.5281/zenodo.16580109}}{10.5281/zenodo.16580109}},
}

@misc{analysis_release,
  author       = {Bennett, Ed and
                  Lucini, Biagio and
                  Mason, David and
                  Piai, Maurizio and
                  Rinaldi, Enrico and
                  Vadacchino, Davide and
                  Zierler, Fabian},
  title        = {Analysis release -- Finite-temperature Yang-Mills theories with the density of states method: towards the continuum limit},
  year         = 2025,
  publisher    = {Zenodo},
  version      = {1.0.0},
  doi          = {\href{{https://doi.org/10.5281/zenodo.16579683}}{10.5281/zenodo.16579683}},
}

@misc{mason_HiRep_LLR_v1.0.0,
  author = {Mason, David and
        Lucini, Biagio and
        Piai, Maurizio and
        Vadacchino, Davide and
        Rinaldi, Enrico},
  title = {First-order phase transitions in {Y}ang-{M}ills
        theories and the density of state method --- {H}i{R}ep
        {LLR} Code v1.0.0},
  month = jul,
  year  = 2023,
  publisher = {Zenodo},
  version = {v1.0.0},
  doi = {\href{{https://doi.org/10.5281/zenodo.8134756}}{10.5281/zenodo.8134756}},
  url= {\url{https://doi.org/10.5281/zenodo.8134756}},
}

@misc{mason_HiRep_LLR_v1.1.0,
  author = {Mason, David and
        Bennett, Ed and
        Lucini, Biagio and
        Piai, Maurizio and
        Vadacchino, Davide and
        Rinaldi, Enrico},
  title = {The density of states method for Symplectic gauge
        theories at finite temperature --- {H}i{R}ep {LLR} Code v1.1.0 },
  month = sep,
  year  = 2024,
  publisher = {Zenodo},
  doi = {\href{https://doi.org/10.5281/zenodo.13807993}{10.5281/zenodo.13807993}},
  url = {\url{https://doi.org/10.5281/zenodo.13807993}},
}

@article{Bennett:2025whm,
    author = "Bennett, Ed and Lucini, Biagio and Mason, David and Piai, Maurizio and Rinaldi, Enrico and Vadacchino, Davide and Zierler, Fabian",
    title = "{Finite-temperature Yang-Mills theories with the density of states method: towards the continuum limit}",
    eprint = "2509.19009",
    archivePrefix = "arXiv",
    primaryClass = "hep-lat",
    reportNumber = "RIKEN-iTHEMS-Report-25",
    month = "9",
    year = "2025"
}

@article{Bennett:2025neg,
    author = "Bennett, Ed",
    title = "{The TELOS Collaboration Approach to Reproducibility and Open Science}",
    eprint = "2504.01876",
    archivePrefix = "arXiv",
    primaryClass = "hep-lat",
    month = "4",
    year = "2025"
}

@article{Panero:2009tv,
    author = "Panero, Marco",
    title = "{Thermodynamics of the QCD plasma and the large-N limit}",
    eprint = "0907.3719",
    archivePrefix = "arXiv",
    primaryClass = "hep-lat",
    doi = "10.1103/PhysRevLett.103.232001",
    journal = "Phys. Rev. Lett.",
    volume = "103",
    pages = "232001",
    year = "2009"
}

@article{Lucini:2012wq,
    author = "Lucini, Biagio and Rago, Antonio and Rinaldi, Enrico",
    title = "{SU($N_c$) gauge theories at deconfinement}",
    eprint = "1202.6684",
    archivePrefix = "arXiv",
    primaryClass = "hep-lat",
    doi = "10.1016/j.physletb.2012.04.070",
    journal = "Phys. Lett. B",
    volume = "712",
    pages = "279--283",
    year = "2012"
}

@article{Lucini:2002ku,
    author = "Lucini, Biagio and Teper, Michael and Wenger, Urs",
    title = "{The Deconfinement transition in SU(N) gauge theories}",
    eprint = "hep-lat/0206029",
    archivePrefix = "arXiv",
    reportNumber = "OUTP-02-28P",
    doi = "10.1016/S0370-2693(02)02556-X",
    journal = "Phys. Lett. B",
    volume = "545",
    pages = "197--206",
    year = "2002"
}

@article{Rindlisbacher:2025dqw,
    author = "Rindlisbacher, Tobias and Rummukainen, Kari and Salami, Ahmed",
    title = "{Confined-deconfined interface tension and latent heat in SU(N) gauge theory}",
    eprint = "2506.15509",
    archivePrefix = "arXiv",
    primaryClass = "hep-lat",
    reportNumber = "HIP-2025-21/TH",
    doi = "10.1103/hxfx-8jqb",
    journal = "Phys. Rev. D",
    volume = "112",
    number = "11",
    pages = "114507",
    year = "2025"
}

@article{Holland:2003kg,
    author = "Holland, K. and Pepe, M. and Wiese, U. J.",
    title = "{The Deconfinement phase transition of Sp(2) and Sp(3) Yang-Mills theories in (2+1)-dimensions and (3+1)-dimensions}",
    eprint = "hep-lat/0312022",
    archivePrefix = "arXiv",
    doi = "10.1016/j.nuclphysb.2004.06.026",
    journal = "Nucl. Phys. B",
    volume = "694",
    pages = "35--58",
    year = "2004"
}

@article{Cossu:2007dk,
    author = "Cossu, Guido and D'Elia, Massimo and Di Giacomo, Adriano and Lucini, Biagio and Pica, Claudio",
    title = "{G(2) gauge theory at finite temperature}",
    eprint = "0709.0669",
    archivePrefix = "arXiv",
    primaryClass = "hep-lat",
    reportNumber = "IFUP-TH-2007-22, BNL-NT-07-36",
    doi = "10.1088/1126-6708/2007/10/100",
    journal = "JHEP",
    volume = "10",
    pages = "100",
    year = "2007"
}

@article{Pepe:2006er,
    author = "Pepe, M. and Wiese, U. -J.",
    title = "{Exceptional Deconfinement in G(2) Gauge Theory}",
    eprint = "hep-lat/0610076",
    archivePrefix = "arXiv",
    doi = "10.1016/j.nuclphysb.2006.12.024",
    journal = "Nucl. Phys. B",
    volume = "768",
    pages = "21--37",
    year = "2007"
}

@article{Huang:2020crf,
    author = "Huang, Wei-Chih and Reichert, Manuel and Sannino, Francesco and Wang, Zhi-Wei",
    title = "{Testing the dark SU(N) Yang-Mills theory confined landscape: From the lattice to gravitational waves}",
    eprint = "2012.11614",
    archivePrefix = "arXiv",
    primaryClass = "hep-ph",
    doi = "10.1103/PhysRevD.104.035005",
    journal = "Phys. Rev. D",
    volume = "104",
    number = "3",
    pages = "035005",
    year = "2021"
}

@article{Halverson:2020xpg,
    author = "Halverson, James and Long, Cody and Maiti, Anindita and Nelson, Brent and Salinas, Gustavo",
    title = "{Gravitational waves from dark Yang-Mills sectors}",
    eprint = "2012.04071",
    archivePrefix = "arXiv",
    primaryClass = "hep-ph",
    doi = "10.1007/JHEP05(2021)154",
    journal = "JHEP",
    volume = "05",
    pages = "154",
    year = "2021"
}

@article{Kang:2021epo,
    author = "Kang, Zhaofeng and Zhu, Jiang and Matsuzaki, Shinya",
    title = "{Dark confinement-deconfinement phase transition: a roadmap from Polyakov loop models to gravitational waves}",
    eprint = "2101.03795",
    archivePrefix = "arXiv",
    primaryClass = "hep-ph",
    doi = "10.1007/JHEP09(2021)060",
    journal = "JHEP",
    volume = "09",
    pages = "060",
    year = "2021"
}

@book{Panico:2015jxa,
    author = "Panico, Giuliano and Wulzer, Andrea",
    title = "{The Composite Nambu-Goldstone Higgs}",
    eprint = "1506.01961",
    archivePrefix = "arXiv",
    primaryClass = "hep-ph",
    reportNumber = "DFPD-2015TH9",
    doi = "10.1007/978-3-319-22617-0",
    publisher = "Springer",
    volume = "913",
    year = "2016"
}

@article{Cirelli:2024ssz,
    author = "Cirelli, Marco and Strumia, Alessandro and Zupan, Jure",
    title = "{Dark Matter}",
    eprint = "2406.01705",
    archivePrefix = "arXiv",
    primaryClass = "hep-ph",
    month = "6",
    year = "2024"
}

@article{Cacciapaglia:2020kgq,
    author = "Cacciapaglia, Giacomo and Pica, Claudio and Sannino, Francesco",
    title = "{Fundamental Composite Dynamics: A Review}",
    eprint = "2002.04914",
    archivePrefix = "arXiv",
    primaryClass = "hep-ph",
    doi = "10.1016/j.physrep.2020.07.002",
    journal = "Phys. Rept.",
    volume = "877",
    pages = "1--70",
    year = "2020"
}

@article{Bennett:2023wjw,
    author = "Bennett, Ed and Holligan, Jack and Hong, Deog Ki and Hsiao, Ho and Lee, Jong-Wan and Lin, C. -J. David and Lucini, Biagio and Mesiti, Michele and Piai, Maurizio and Vadacchino, Davide",
    title = "{Sp(2N) Lattice Gauge Theories and Extensions of the Standard Model of Particle Physics}",
    eprint = "2304.01070",
    archivePrefix = "arXiv",
    primaryClass = "hep-lat",
    reportNumber = "CTPU-PTC-23-09, PNUTP-23/A02",
    doi = "10.3390/universe9050236",
    journal = "Universe",
    volume = "9",
    number = "5",
    pages = "236",
    year = "2023"
}

@article{Witten:1984rs,
    author = "Witten, Edward",
    title = "{Cosmic Separation of Phases}",
    reportNumber = "PRINT-84-0400 (IAS,PRINCETON)",
    doi = "10.1103/PhysRevD.30.272",
    journal = "Phys. Rev. D",
    volume = "30",
    pages = "272--285",
    year = "1984"
}

@article{ET:2019dnz,
    author = "Maggiore, Michele and others",
    collaboration = "ET",
    title = "{Science Case for the Einstein Telescope}",
    eprint = "1912.02622",
    archivePrefix = "arXiv",
    primaryClass = "astro-ph.CO",
    doi = "10.1088/1475-7516/2020/03/050",
    journal = "JCAP",
    volume = "03",
    pages = "050",
    year = "2020"
}

@article{Caprini:2019egz,
    author = "Caprini, Chiara and others",
    title = "{Detecting gravitational waves from cosmological phase transitions with LISA: an update}",
    eprint = "1910.13125",
    archivePrefix = "arXiv",
    primaryClass = "astro-ph.CO",
    reportNumber = "DESY-19-159, IPPP/19/27, HIP-2019-14/TH, MITP/19-066, IFT-UAM/CSIC-19-139",
    doi = "10.1088/1475-7516/2020/03/024",
    journal = "JCAP",
    volume = "03",
    pages = "024",
    year = "2020"
}

@article{Borsanyi:2022xml,
    author = "Borsanyi, S. and R. Kara and Fodor, Z. and Godzieba, D. A. and Parotto, P. and Sexty, D.",
    title = "{Precision study of the continuum SU(3) Yang-Mills theory: How to use parallel tempering to improve on supercritical slowing down for first order phase transitions}",
    eprint = "2202.05234",
    archivePrefix = "arXiv",
    primaryClass = "hep-lat",
    doi = "10.1103/PhysRevD.105.074513",
    journal = "Phys. Rev. D",
    volume = "105",
    number = "7",
    pages = "074513",
    year = "2022"
}

@article{Pasechnik:2023hwv,
    author = "Pasechnik, Roman and Reichert, Manuel and Sannino, Francesco and Wang, Zhi-Wei",
    title = "{Gravitational waves from composite dark sectors}",
    eprint = "2309.16755",
    archivePrefix = "arXiv",
    primaryClass = "hep-ph",
    doi = "10.1007/JHEP02(2024)159",
    journal = "JHEP",
    volume = "02",
    pages = "159",
    year = "2024"
}

@article{Reichert:2021cvs,
    author = "Reichert, Manuel and Sannino, Francesco and Wang, Zhi-Wei and Zhang, Chen",
    title = "{Dark confinement and chiral phase transitions: gravitational waves vs matter representations}",
    eprint = "2109.11552",
    archivePrefix = "arXiv",
    primaryClass = "hep-ph",
    doi = "10.1007/JHEP01(2022)003",
    journal = "JHEP",
    volume = "01",
    pages = "003",
    year = "2022"
}

@article{Langfeld:2012ah,
    author = "Langfeld, Kurt and Lucini, Biagio and Rago, Antonio",
    title = "{The density of states in gauge theories}",
    eprint = "1204.3243",
    archivePrefix = "arXiv",
    primaryClass = "hep-lat",
    doi = "10.1103/PhysRevLett.109.111601",
    journal = "Phys. Rev. Lett.",
    volume = "109",
    pages = "111601",
    year = "2012"
}

@article{Langfeld:2015fua,
    author = "Langfeld, Kurt and Lucini, Biagio and Pellegrini, Roberto and Rago, Antonio",
    title = "{An efficient algorithm for numerical computations of continuous densities of states}",
    eprint = "1509.08391",
    archivePrefix = "arXiv",
    primaryClass = "hep-lat",
    doi = "10.1140/epjc/s10052-016-4142-5",
    journal = "Eur. Phys. J. C",
    volume = "76",
    number = "6",
    pages = "306",
    year = "2016"
}

@article{Lucini:2023irm,
    author = "Lucini, Biagio and Mason, David and Piai, Maurizio and Rinaldi, Enrico and Vadacchino, Davide",
    title = "{First-order phase transitions in Yang-Mills theories and the density of state method}",
    eprint = "2305.07463",
    archivePrefix = "arXiv",
    primaryClass = "hep-lat",
    reportNumber = "RIKEN-iTHEMS-Report-23 ET-0164A-23",
    doi = "10.1103/PhysRevD.108.074517",
    journal = "Phys. Rev. D",
    volume = "108",
    number = "7",
    pages = "074517",
    year = "2023"
}

@article{Bennett:2024bhy,
    author = "Bennett, Ed and Lucini, Biagio and Mason, David and Piai, Maurizio and Rinaldi, Enrico and Vadacchino, Davide",
    title = "{Density of states method for symplectic gauge theories at finite temperature}",
    eprint = "2409.19426",
    archivePrefix = "arXiv",
    primaryClass = "hep-lat",
    reportNumber = "RIKEN-iTHEMS-Report-24 ET-0515A-24, RIKEN-iTHEMS-Report-24, ET-0515A-24",
    doi = "10.1103/PhysRevD.111.114511",
    journal = "Phys. Rev. D",
    volume = "111",
    number = "11",
    pages = "114511",
    year = "2025"
}

@article{DelDebbio:2008zf,
    author = "Del Debbio, Luigi and Patella, Agostino and Pica, Claudio",
    title = "{Higher representations on the lattice: Numerical simulations. SU(2) with adjoint fermions}",
    eprint = "0805.2058",
    archivePrefix = "arXiv",
    primaryClass = "hep-lat",
    reportNumber = "NI08019, BNL-NT-08-15, CP3-ORIGINS-2010-15",
    doi = "10.1103/PhysRevD.81.094503",
    journal = "Phys. Rev. D",
    volume = "81",
    pages = "094503",
    year = "2010"
}
\end{document}